\documentclass[aps,nofootinbib]{revtex4}

\usepackage{amsmath,amssymb,graphicx,float}
\usepackage{color}
\usepackage[normalem]{ulem}

\newcommand{\mathsym}[1]{{}}
\newcommand{\unicode}[1]{{}}

\begin{document}

\title{Role of correlations in the maximum distribution of multiscale stationary Markovian processes}
\author{S. Miccich\`e}
\affiliation{Dipartimento di Fisica e Chimica - Emilio Segr\`e, Universit\`a degli Studi di Palermo, Viale delle Scienze, Ed. 18, 90128, Palermo, Italy}
                
\date{\today}

\begin{abstract}

We are interested in investigating the statistical properties of extreme values for strongly correlated variables.  The starting motivation is to  understand how the strong-correlation properties of power-law distributed processes affect the possibility of exploring the whole domain of a stochastic process (the real axis in most cases) when performing time-average numerical simulations and how this relates to the numerical evaluation of the autocorrelation function. 

We  show that correlations decrease the heterogeneity of the maximum values. Specifically, through numerical simulations we observe that for strongly correlated variables whose probability distribution function decays like a power-law $1/x^\alpha$, the maximum distribution has a tail compatible with a $1/x^{\alpha+2}$ decay, while for i.i.d. variables we expect a $1/x^\alpha$ decay. As a consequence, we also show that the numerically estimated  autocorrelation function converges to the theoretical prediction according to a factor that depends on the length of the simulated time-series $n$ according to a power-law: $1/n^{\alpha^\delta}$ with $\delta<1$, This accounts for a very slow convergence rate.

\end{abstract}

\maketitle

\section{Introduction} \label{intro}

Extreme value theory \cite{embrechts} is a research field that investigates a crucial issue in many disciplines ranging from finance \cite{markowitz, taqqu} to natural and social phenomena \cite{albeverio} to disordered systems in physics \cite{aging}. In a nutshell, the problem is to investigate the statistics of the maximum (or the minimum) of a set of stochastic variables. In the case when such variables are independent and identically distributed (i.i.d.) many analytical results are known \cite{maxdistrep}. Indeed, as long as the variables are weakly correlated, the results obtained for the i.i.d. case can be easily generalized. When the variables are strongly correlated little is known.

We are interested in investigating the statistical properties of extreme values for strongly correlated variables. Moreover, we will mainly focus on stationary processes that are also power-law distributed. The starting motivation is to understand how the strong-correlation properties of power-law distributed processes affect the possibility of exploring the whole domain of a stochastic process (the real axis in most cases) when performing time-average numerical simulations and how this relates to the numerical evaluation of the autocorrelation function of a stochastic process. We have already shown \cite{ACnewpaper} how  the simulated autocorrelation function $R_{\rm{L}}(\tau)$ converges to the predicted autocorrelation function $R(\tau)$ at a rate that essentially depends by the pdf tail computed in $L$, where $L$ is the maximum values that is attained when numerically simulating the stochastic process. As such, here we are interested in understanding how $L$ is related to the length $n$ of the simulated time-series.

Without pretending to be too general, we will here consider  stationary Markovian stochastic processes that can be described by a nonlinear Langevin equation, with a white-noise term, and that admit a Fokker-Planck equation \cite{risken, gardiner}, as in Refs. \cite{myuno, mydue, mytre}. We will leave for a future work the generalization to other classes of stochastic processes.

We will show that correlations modify the average maximum and the maximum distribution expected for i.i.d. variables in the direction of decreasing the heterogeneity of the maximum distribution. Specifically, through numerical simulations we observe that for strongly correlated power-law distributed processes the maximum distribution has a tail compatible with a $1/x^{\alpha+2}$ decay, while for i.i.d. variables we expect a $1/x^\alpha$ decay, where $\alpha$ is the exponent of the power-law decaying probability distribution function (pdf). As a consequence, we also show that the {\em{error}} that one makes when numerically evaluating the autocorrelation function $R_{\rm{L}}(\tau)$ depends on $n$ according to a power-law: $1/n^{\alpha^\delta}$ with $\delta<1$, This accounts for a very slow convergence rate.

The paper is organized as follows: in section \ref{maxav} we investigate the statistical properties of the average maximum for strongly correlated variables. In section \ref{maxdist} we investigate the maximum distribution for strongly correlated variables. We discuss our results and draw our conclusions in section \ref{concl}.
 
\section{Average maximum of correlated power-law distributed processes} \label{maxav}

The importance of power-law distributed processes has been throroughtly elucidated in Ref. \cite{newman}, that, despite its publication year, can be still considered as {\em{the}} reference paper for anyone interested in accessing the power-law world.

In particular, in section III.C the author shows a nice argument for obtaining the average maximum value as a function of the length $n$ of the considered time-series:
\begin{eqnarray}
                             \langle x_{\rm{max}} \rangle \approx n^{1/(\alpha-1)} \label{expiid}
\end{eqnarray}	
where $\alpha$ is the exponent of the probability distribution function (pdf) power-law tail: $p(x) \approx 1/x^\alpha$.

\begin{figure}[H]
\begin{center}
                      {\includegraphics[scale=0.30]{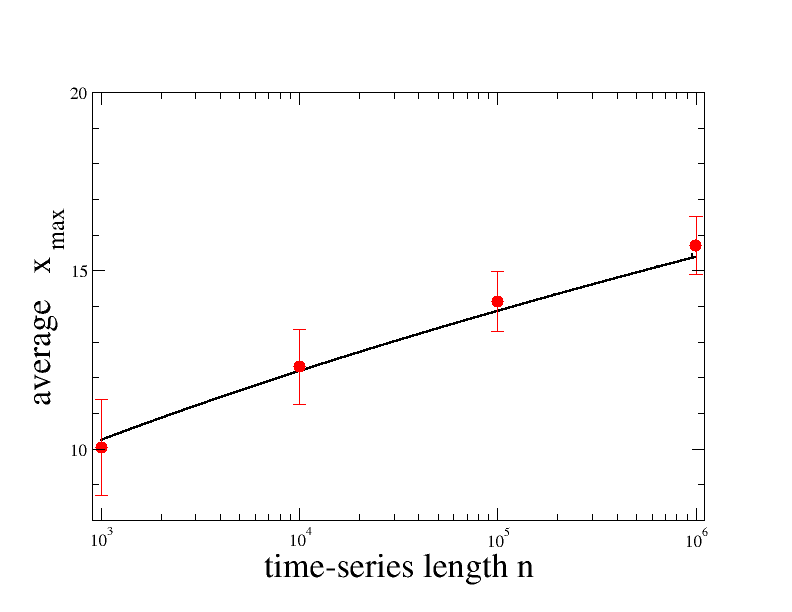}}
\end{center}
\caption{{\sl{Average maximum $ \langle x_{\rm{max}} \rangle$ for an Ornstein-Uhlembeck process with $\gamma=0.1$. The red circles show the results of numerical simulations while the black solid lines show the theoretical results expected according to the procedure of Ref. \cite{newman} section III.C. Each circle represents the median and standard deviation computed over $M=10^6$ simulations.}}}\label{OU-GAUSSiid-POWiid}
\end{figure}

Indeed, the procedure sketched in Ref. \cite{newman} section 	III.C is quite general and can be also applied to processes that have a pdf with tails different from a power-law. In fact in Fig. \ref{OU-GAUSSiid-POWiid} we show $ \langle x_{\rm{max}} \rangle$ as a function of $n$, for 
an Ornstein-Uhlembeck process \cite{OU} with unitary diffusion coefficient and drift coefficient $h(x)=-\gamma \, x$ whose pdf is given by:
\begin{eqnarray}
                           p_{\rm{OU}}(x)={ \sqrt{\gamma} \over \sqrt{2 \pi}} \, \exp^{-\gamma/2 x^2}.                                         
\end{eqnarray}
The red circles show the results of numerical simulations while the black solid lines show the theoretical results expected according to the procedure of Ref. \cite{newman} section III.C. Each circle represents the median and standard deviation computed over $M=10^6$ simulations. 
The required integrations have been performed only numerically. 
The agreement between numerical simulations and theoretical prediction is quite good.

\begin{figure}[H]
\begin{center}
                      {\includegraphics[scale=0.30]{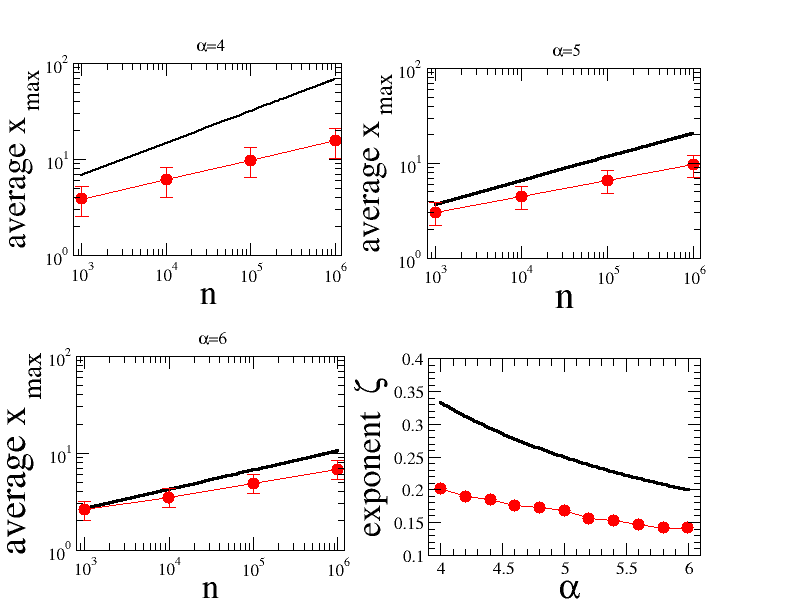}}
\end{center}
\caption{{\sl{Average maximum $ \langle x_{\rm{max}} \rangle$ for the process of Eq. \ref{chimera} with $\alpha=4$ (top-left panel), $\alpha=5$ (top-right panel) and $\alpha=6$ (bottom-left panel). The red circles show the results of numerical simulations while the black solid lines show the theoretical results expected according to the procedure of Ref. \cite{newman} section III.C. Each circle represents the median and standard deviation computed over $M=10^6$ simulations. In the bottom-right panel we show the fitted exponents $\zeta$ as a function of $\alpha$.}}} \label{CHI}
\end{figure}

Let us now consider a stationary Markovian stochastic process described by a nonlinear Langevin equation with unitary diffusion coefficient and drift coefficient \cite{myuno,mydue,mytre,ACnewpaper}:
\begin{eqnarray}
          &&               h(x)= \left \{ \begin{array}{cc}
               -2 \sqrt{V_0} \tan (\sqrt{V_0}x) &{\rm{if}}~~|x| \leqslant L ,\\
                 &   \\
                 (1-\sqrt{1+4~V_1})/ x   &{\rm{if}}~~|x| >   L .   
                                  \end{array} \right. \label{chimera} \\
          &&  V_1=L \tan\bigl( \sqrt{V_0} L\bigl) \Bigl(1+L \tan\bigl( \sqrt{V_0} L\bigl)\Bigl)   \nonumber 
\end{eqnarray} 
This process admits a stationary pdf with power-law tails $p(x) \approx x^{-\alpha}$ and a power-law decaying autocorrelation function $\rho(\tau) \approx \tau^{-\beta}$ with $\beta=(\alpha-3)/2$. When $\alpha \in]3,5[$ the process is long-range correlated. When $\alpha>5$ the process is short-range correlated. When $\alpha \le 3$ the variance of the process is not defined. Hereafter we will always consider $\alpha>3$. In Fig. \ref{CHI} we show $\langle x_{\rm{max}}\rangle$ as a function of $n$, for $\alpha=4$ (top-left panel), $\alpha=5$ (top-right panel) and $\alpha=6$ (bottom-left). The red circles show the results of numerical simulations while the black solid lines show the theoretical results expected according to the procedure of Ref. \cite{newman} section III.C. Each circle represents the median and standard deviation computed over $M=10^6$ simulations. Numerical simulations are quite different from predictions. In fact,  the average maximum still grows like a power-law $\langle x_{\rm{max}} \rangle \approx n^\zeta$, but with an exponent $\zeta$ much smaller than the one predicted in Ref. \cite{newman}, see Eq. \ref{expiid}. In the bottom-right panel we show the fitted exponents as a function of $\alpha$. The disagreement with the theoretical prediction is quite evident.

Indeed we believe that this result clearly indicates that processes correlations play a central role in determining the properties of the maximum distribution and, in particular, of $\langle x_{\rm{max}}\rangle$.

\section{Maximum distribution of correlated processes} \label{maxdist}

Hereafter, we would like to show how correlations modify the properties of the maximum distribution for which exact results are known in the case of i.i.d. stochastic variables Ref. \cite{maxdistrep}. In fact, one main result is that for i.i.d. gaussian distributed stochastic variables the maximum distribution is the Gumbel-type distribution \cite{gumbel} while for i.i.d. power-law distributed variables we have a Frechet-type distribution \cite{frechet, fisher}. In particular, by following the lines of section 3 in Ref. \cite{maxdistrep} for i.i.d. processes one gets:
\begin{eqnarray}
                          &&  p_{\rm{gauss}}(x)={1 \over \sqrt{2 \pi} \sigma} \, \exp^{-{x^2 \over 2 \sigma^2}}  \mapsto G(X)= {n \over \sqrt{2 \pi} \sigma} \, \exp^{-{X^2 \over 2 \sigma^2}} \, 
                                              \Bigl [ 1-{1 \over 2} \, {\rm{Erf}}\Bigl({X \over \sqrt{2} \sigma}\Bigr)\Bigl]^n \label{gumbel} \\
                          &&  p_{\rm{pow}}(x) \approx A_0 \, {1 \over x^\alpha}  \quad \quad \quad \quad  \, \,   \mapsto F(X) \approx A_0 \, (\alpha-1) \, n \, {1 \over X^\alpha} \label{frechet}
\end{eqnarray}
where, in the case of power-law distributed processes, $A_0$ is an appropriate normalization constant and we consider the asymptotic behaviour only. 

\begin{figure}[H]
\begin{center}
                      {\includegraphics[scale=0.30]{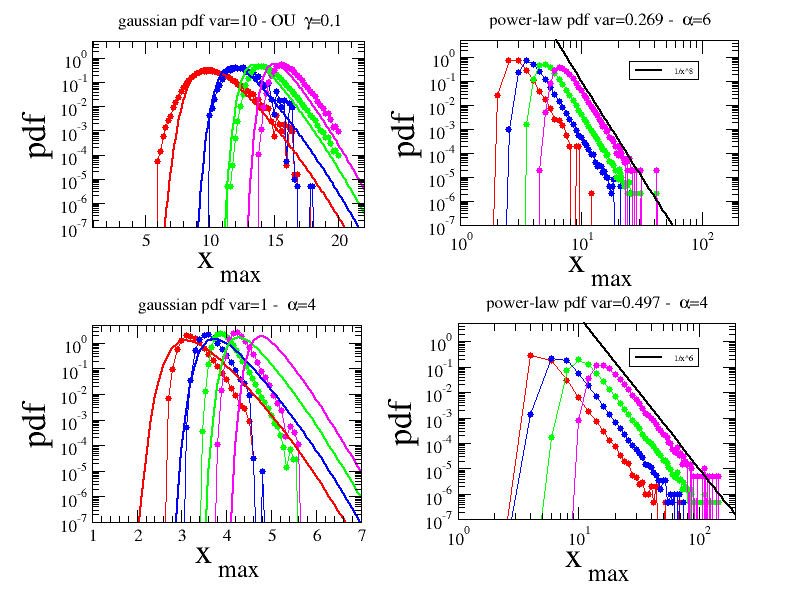}}
\end{center}
\caption{{\sl{Maximum distribution for an Ornstein-Uhlembeck process with $\gamma=0.1$ (top-left panel), the process of Eq. \ref{chimera} with $\alpha=6$ (top-right panel), the process of Eq. \ref{chimera} with $\alpha=4$ (bottom-right panel) and a power-law correlated process normally distributed (bottom-left panel). The solid lines in the two left panels show the $G(x)$ function of Eq. \ref{gumbel} at different values of $n$. The black solid lines in the right panels show a reference power-law function with exponent indicated in the legends. The empirical distributions are computed over $M=10^6$ realizations. Different colours refer to simulations with different values of $n$: Red for $n=10^3$,  Blue for $n=10^4$, Green for $n=10^5$, Magenta for $n=10^6$,}}}\label{MAXdistr}
\end{figure}

In the top-left panel of Fig. \ref{MAXdistr} we show $G(x)$ (solid lines) for different values of $n$ together with the empirical distributions obtained trough numerical simulations of an Ornstein-Uhlembeck process with $\gamma=0.1$. We have already seen in the previous section that there is a good agreement between theoretical predictions and numerical simulations for $\langle x_{max} \rangle$. However, at the level of the maximum distribution one can notice that in the OU process $x_{max}$ values seem closer to $\langle x_{max} \rangle$ than for the i.i.d. case. Somehow, the effect of correlations is that of introducing a sort of viscosity that bound the  $x_{max}$ values to their averages.
Such an effect is even mmore pronounced in the case of the power-law correlated power-law distributed process of Eq. \ref{chimera}.  The two right panels of Fig. \ref{MAXdistr} show, for different values of $n$, the empirical maximum distributions obtained trough numerical simulations of the process of Eq. \ref{chimera} with $\alpha=6$ (top-right panel) and $\alpha=4$ (bottom-right panel). We see that in this case the tails of the maximum distribution are compatible with a power-law decay $1/X^{\alpha+2}$, while for i.i.d. processes we expect a decay $F(X) \approx 1/X^\alpha$. All distributions have been computed over $M=10^6$ different realizations of the process, each realization having length $n$.

In order to better elucidate the role of correlation in shaping the maximum distribution, we have considered a coordinate transformation $x \mapsto y$ that starting from the process of Eq. \ref{chimera} leads to a gaussian process with null average and unitary variance in the $y$ domain, see section III of Ref. \cite{myuno} for more details. We have shown in Ref. \cite{myuno} and Ref. \cite{mydue} that this is a power-law correlated gaussian process. In the bottom-left panel of Fig. \ref{MAXdistr} we show $G(x)$ (solid lines) for different values of $n$ together with the empirical distributions obtained trough numerical simulations of such process. Since the process is gaussian, we do not observe power-law tails in the maximum distribution. However, the simulated distributions are much narrower than $G(x)$, thus confirming the viscosity effect mentioned above. Indeed, the effect is much more pronounced here than for the OU process, which again can be considered as a signature of the role of strong correlations in shaping the maximum distribution.

\begin{figure}[H]
\begin{center}
                      {\includegraphics[scale=0.30]{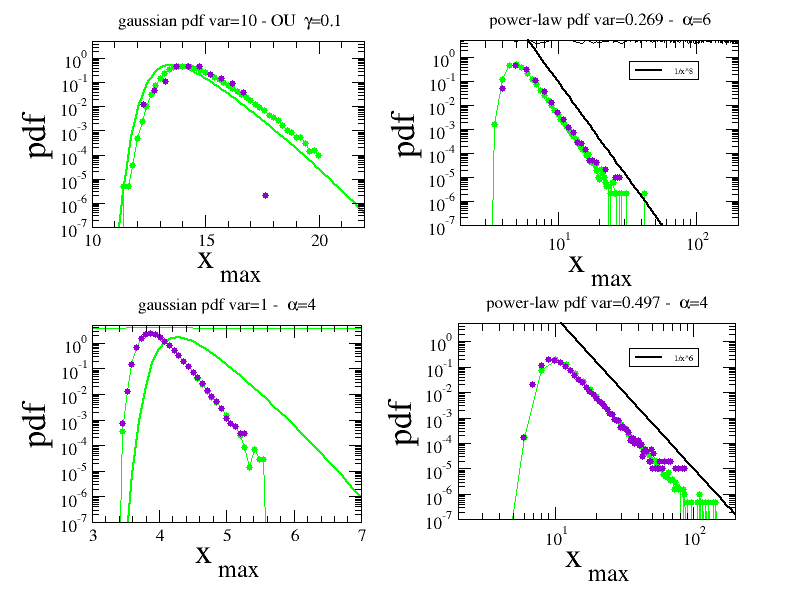}
                       \includegraphics[scale=0.30]{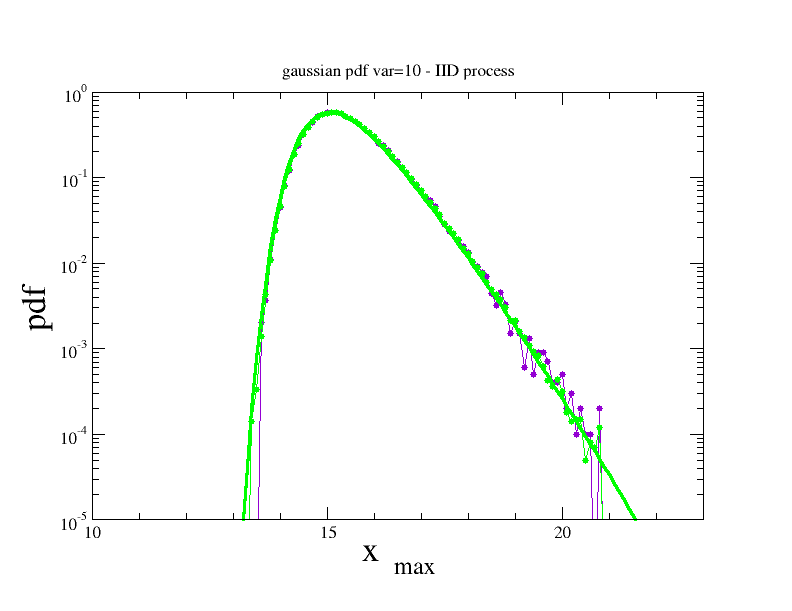}}
\end{center}
\caption{{\sl{{\bf{Left panels}} - Maximum distribution for an Ornstein-Uhlembeck process with $\gamma=0.1$ (top-left panel), the process of Eq. \ref{chimera} with $\alpha=6$ (top-right panel), the process of Eq. \ref{chimera} with $\alpha=4$ (bottom-right panel) and a  power-law correlated process normally distributed (bottom-left panel). The solid lines in the two left panels show the $G(x)$ function of Eq. \ref{gumbel} for $n=10^5$. The black solid lines in the right panels show a reference power-law function with exponent indicated in the legends. {\bf{Right panel}} - Maximum distribution for a gaussian i.i.d. process.  The solid line shows $G(x)$ of Eq. \ref{gumbel} for $n=10^6$. In all panels green circles refer to simulations over $M=10^6$ realizations and violet circles refer to simulations done in $10$ bunches each one composed of $M'=10^5$ realizations. The length of each simulation is $n=10^5$ in the left panels and $n=10^6$ in the right panel.}}}\label{MAXdistrslow}
\end{figure}

The simulations with $n=10^3$, $n=10^4$ and $n=10^5$ shown in the above figures have all been obtained with the same seed and the same starting point $x_0=0.1$. At the end of each of the $M$ realizations, the final point of the process' realization was used as the starting point of the realization of the next iteration. In the case of $n=10^6$, in order to reduce the computational time needed to perform such simulations, we parallelized the computation by splitting the $M$ realizations into $10$ bunches each one composed of $M'=10^5$ realizations of length $n=10^6$. In order to better illustrate the effect of this choice, in the four left panels of Fig. \ref{MAXdistrslow}, for the same processes as in Fig. \ref{MAXdistr}, we show the maximum distribution for $n=10^5$ when we perform a unique simulation (green circles) and when we split the simulation in $10$ bunches each one composed of $M'=10^5$ realizations of length $n=10^5$. In the case when we perform a unique simulation we can attain much larger maximum values. This is again a signature of the fact that the more a process is correlated the smaller the possibility of attaining large values. In fact, in the right panel of Fig. \ref{MAXdistrslow} we show the same quantities for an i.i.d. gaussian distributed process: when we perform a unique simulation of $M=10^6$ realizations each one of length $n=10^6$ maxima are distributed in the interval $[13.31, 20.73]$ while when we split the simulation in $M'=10$ bunches of $M=10^5$ realizations each one of length $n=10^6$ maxima are distributed in the interval $[13.50, 20.72]$

In summary, we might speculate that the effect of correlations is that of bounding the  $x_{max}$ values to their average values, although probably the type of distribution is of the same as for  i.i.d. variables.

\section{Conclusions} \label{concl}

The viscosity effect mentioned above has already been observed in Ref. \cite{mydue} and Ref. \cite{mytre} where the mean first passage time and the first passage time distribution where respectively investigated. In fact, we had shown that the more the process is correlated the larger the average time $T_x(\Lambda)$ needed to reach a certain boundary $\Lambda$, thus indicating that correlations enhance the time spent in a given position and therefore diminishes the likelihood that it reaches nearby positions by chance.

The results shown above, are also relevant for determining the error that one makes when numerically evaluating the autocorrelation function due to the fact that any simulated time series ${\cal{X}}=\{x_1, x_2, \cdots x_n\}$ is bounded, i.e. $|x_i| \le L$, which in turn is a function of $n$. In fact, in Ref. \cite{ACnewpaper} we have shown that the simulated autocorrelation function $R_{\rm{L}}(\tau)$ converges to the predicted autocorrelation function $R(\tau)$ at a rate that essentially depends by the pdf tail computed in $L$. Clearly $\langle x_{\rm{max}} \rangle$ is a very reasonable proxy for $L$. In the case of the process of Eq. \ref{chimera}, $R_{\rm{L}}(\tau) \approx L^{1-\alpha/2} \, R(\tau)$  \cite{ACnewpaper}. We have shown above that $\langle x_{\rm{max}}\rangle \approx n^\beta$ where $\beta$ depends on $\alpha$ according to the empirical law illustrated in the bottom-right panel of Fig. \ref{CHI}. As such the simulated autocorrelation function converges to the predicted autocorrelation function $R(\tau)$ at a rate that is given by $1/n^{\beta (\alpha/2-1)}$. By inspecting the bottom-right panel of Fig. \ref{CHI} one may argue that $\beta$ is such that the convergence rate is of the order of  $1/n^{\alpha^\delta}$ with $\delta<1$, i.e. a very slow convergence rate. This indicates that by increasing the length of the time series the advantage in simulating the autocorrelation function is small.

The investigation of the way the maximum distribution depends on $n$ gives us indication about the way the process domain is explored during the time evolution. This is particularly relevant when performing numerical simulations as the capability of attaining large values in often crucial in the estimation of many statistical quantities such as autocorrelation function of power-law tails that are relevant for out understanding of the process we are considering. However, it is worth mentioning that the topics we are addressing here are related but different from the issues related to ergodicity \cite{aging,ergo,flandoli}. In fact, we are not giving here any prescription about the fact that the process be or not be ergodic. We are rather giving a few technical indications about the effectiveness of time-average numerical simulations, which is preliminary to the investigation of the ergodic properties of a certain process. 

Despite the specific motivation that inspired this work, we believe that our results are general enough to be of relevance in many disparate research fields. On our side we leave for future work the generalization of these results to other classes of stochastic processes as well as the investigation of the role of correlation properties in aging phenomena in physical and non physical complex systems.
 



\bibliographystyle{IEEEtran}
%

\end{document}